\documentclass[11pt]{article}
\usepackage{amsmath}
\usepackage{amsfonts}
\usepackage{amssymb}
\usepackage{graphicx}

\newcommand{\Title}{\bf \sf \huge \noindent}
\newcommand{\Author}{\bf \sf \large \noindent}
\begin{document}
\mbox{  } \vskip 1.2cm

\Title{Intermediate-statistics spin waves}

\noindent \Huge
------------------------------------------------------
\vskip 0.5cm
\Author{Wu-Sheng Dai and Mi Xie}

{\it \noindent \small Department of Physics, Tianjin University,
Tianjin 300072, P.
R. China}\\
{\it \noindent \small LiuHui Center for Applied Mathematics, Nankai
University \& Tianjin
University, Tianjin 300072, P. R. China}\\
{\it \noindent \small Email: \rm daiwusheng@tju.edu.cn,
xiemi@tju.edu.cn}

\thispagestyle{myheadings} \markboth{Preprint}{\underline{J. Stat.
Mech. (2009) P04021}} \vskip 1cm \rm \normalsize

\noindent {\bf Abstract:} In this paper, we show that spin waves,
the elementary excitation of the Heisenberg magnetic system, obey a
kind of intermediate statistics with a finite maximum occupation
number $n$. We construct an operator realization for the
intermediate statistics obeyed by magnons, the quantized spin waves,
and then construct a corresponding intermediate-statistics
realization for the angular momentum algebra in terms of the
creation and annihilation operators of the magnons. In other words,
instead of the Holstein-Primakoff representation, a bosonic
representation subject to a constraint on the occupation number, we
present an intermediate-statistics representation with no
constraints. In this realization, the maximum occupation number is
naturally embodied in the commutation relation of creation and
annihilation operators, while the Holstein-Primakoff representation
is a bosonic operator relation with an additional putting-in-by-hand
restriction on the occupation number. We deduce the
intermediate-statistics distribution function for magnons from the
intermediate-statistics commutation relation of the creation and
annihilation operators directly, which is a modified Bose-Einstein
distribution. Based on these results, we calculate the dispersion
relations for ferromagnetic and antiferromagnetic spin waves. The
relations between the intermediate statistics that magnons obey and
the other two important kinds of intermediate statistics, Haldane-Wu
statistics and the fractional statistics of anyons, are discussed.
We also compare the spectrum of the intermediate-statistics spin
wave with the exact solution of the one-dimensional $s=1/2$
Heisenberg model, which is obtained by the Bethe ansatz method. For
ferromagnets, we take the contributions from the interaction between
magnons (the quartic contribution), the next-to-nearest-neighbor
interaction, and the dipolar interaction into account for comparison
with the experiment.

\noindent

\newpage
\vskip 1cm \noindent
---------------------------------------------------------------------------------------------------------------------
\tableofcontents

\vskip 1cm \noindent
---------------------------------------------------------------------------------------------------------------------

\section{Introduction}

At low temperatures, the elementary excitations of a magnetic
system, a periodic system of localized spins coupled by exchange
interaction, are magnons, the quantized spin waves. In this paper,
we show that magnons obey a kind of intermediate statistics in which
the maximum number of particles in any quantum state is neither $1$
nor $\infty$, but equals a finite number $n$. That is to say,
magnons are intermediate-statistics type quasiparticles.

Let us first recall the common treatment of a magnetic system. Take
the ferromagnet as an example. The Hamiltonian of the Heisenberg
model of a ferromagnetic system describing the exchange interaction
between neighboring
spins reads%
\begin{equation}
H=\sum_{mn}J_{mn}\mathbf{S}_{m}\cdot\mathbf{S}_{n},\label{Hmg}%
\end{equation}
where $J_{mn}$ is the exchange coefficient and $\mathbf{S}_{m}$ and
$\mathbf{S}_{n}$ are the spins at $m$-th and $n$-th sites. For such
a spin system, introducing the spin deviation operator at site
$\mathbf{\ell}$,
\begin{equation}
N_{\mathbf{\ell}}=S-S_{\mathbf{\ell}}^{z},
\end{equation}
where $S_{\mathbf{\ell}}^{z}$ is the $z$-component of the spin
operator $\mathbf{S}_{\mathbf{\ell}}$ and
$\mathbf{S}_{\mathbf{\ell}}^{2}=S\left(
S+1\right)  $, we have%
\begin{align}
S_{\mathbf{\ell}}^{+}\left\vert N_{\mathbf{\ell}}\right\rangle  &
=\sqrt{2S-\left(  N_{\mathbf{\ell}}-1\right)  }\sqrt{N_{\mathbf{\ell}}%
}\left\vert N_{\mathbf{\ell}}-1\right\rangle ,\nonumber\\
S_{\mathbf{\ell}}^{-}\left\vert N_{\mathbf{\ell}}\right\rangle  &
=\sqrt{N_{\mathbf{\ell}}+1}\sqrt{2S-N_{\mathbf{\ell}}}\left\vert
N_{\mathbf{\ell}}+1\right\rangle ,\nonumber\\
S_{\mathbf{\ell}}^{z}\left\vert N_{\mathbf{\ell}}\right\rangle  &
=S-N_{\mathbf{\ell}},\label{spin}%
\end{align}
where $\left\vert N_{\mathbf{\ell}}\right\rangle $ is the eigenstate
of the
spin deviation operator, i.e., $N_{\mathbf{\ell}}\left\vert N_{\mathbf{\ell}%
}\right\rangle =N_{\mathbf{\ell}}\left\vert
N_{\mathbf{\ell}}\right\rangle $, $\left[
S_{\mathbf{\ell}}^{+},S_{\mathbf{\ell}^{\prime}}^{-}\right]
=2S_{\mathbf{\ell}}^{z}\delta_{\mathbf{\ell\ell}^{\prime}}$, and
$\left[
S_{\mathbf{\ell}}^{z},S_{\mathbf{\ell}^{\prime}}^{\pm}\right]  =\pm
S_{\mathbf{\ell}}^{\pm}\delta_{\mathbf{\ell\ell}^{\prime}}$
\cite{GP}. A natural restriction is
\begin{equation}
0\leq N_{\mathbf{\ell}}\leq2S,\label{NltS}%
\end{equation}
since $S_{\mathbf{\ell}}^{z}$ must be less than the magnitude of
$\mathbf{S}_{\mathbf{\ell}}$. The relation (\ref{spin}) leads to an
operator
realization for the angular momentum:%
\begin{align}
S_{\mathbf{\ell}}^{+}  &  =\sqrt{2S-N_{\mathbf{\ell}}}a_{\mathbf{\ell}}%
^{Bose},\nonumber\\
S_{\mathbf{\ell}}^{-}  &  =a_{\mathbf{\ell}}^{Bose\dagger}\sqrt
{2S-N_{\mathbf{\ell}}},\nonumber\\
S_{\mathbf{\ell}}^{z}  &  =S-N_{\mathbf{\ell}},\label{HP}%
\end{align}
where $a_{\mathbf{\ell}}^{Bose}$, $a_{\mathbf{\ell}}^{Bose\dagger}$,
and $N_{\mathbf{\ell}}^{Bose}$ satisfy the bosonic commutation
relation:
\begin{align}
\left[
a_{\mathbf{\ell}}^{Bose},a_{\mathbf{\ell}^{\prime}}^{Bose\dagger
}\right]   &  =\delta_{\mathbf{\ell\ell}^{\prime}},\nonumber\\
\left[
N_{\mathbf{\ell}},a_{\mathbf{\ell}^{\prime}}^{Bose\dagger}\right]
&
=a_{\mathbf{\ell}}^{Bose\dagger}\delta_{\mathbf{\ell\ell}^{\prime}}\text{\
and }\left[
N_{\mathbf{\ell}},a_{\mathbf{\ell}^{\prime}}^{Bose}\right]
=-a_{\mathbf{\ell}}^{Bose}\delta_{\mathbf{\ell\ell}^{\prime}}.\text{\ }%
\label{BEc}%
\end{align}
From equation (\ref{BEc}), we can see that
$a_{\mathbf{\ell}}^{Bose\dagger}$ creates and
$a_{\mathbf{\ell}}^{Bose}$ annihilates a localized spin deviation at
a single site. This is the Holstein-Primakoff representation of
angular momentum algebra \cite{HP}.

It is known that the Holstein-Primakoff representation is not a
genuine bosonic realization of angular momentum algebra, since
though the operators satisfy the bosonic commutation relation
(\ref{BEc}), there still exists an additional restriction on the
value of the spin deviation $N_{\mathbf{\ell}}$, equation
(\ref{NltS}). When $N_{\mathbf{\ell}}>2S$, the representation is not
faithful, while in the Bose-Einstein case, $N_{\mathbf{\ell}}$ can
take on any value. That is to say, the Holstein-Primakoff
representation corresponds essentially to a kind of intermediate
statistics with a maximum occupation number $2S$. In the
Holstein-Primakoff representation, though there exists a maximum
occupation number $n=2S$, the maximum occupation number is not
embodied in the operator relation (\ref{BEc}); in fact, the
Holstein-Primakoff representation is a bosonic realization with a
putting-in-by-hand restriction on the occupation number. As a
result, however, when using the Holstein-Primakoff representation to
solve the spectrum, only the operator relation is used, so the
influence of the restriction on the occupation number is ignored.

The above analysis shows that the spin deviation does not obey
Bose-Einstein statistics, but obeys a kind of intermediate
statistics with a finite maximum occupation number. In fact, as
discussed in \cite{OurPhysica,OurPRE}, if one wants to construct an
operator realization for the angular momentum algebra by a single
set of creation and annihilation operators, he needs a kind of
intermediate statistics rather than the Bose-Einstein or Fermi-Dirac
case. Nevertheless, in the common treatment of spin waves, the spin
deviations are regarded as bosons: the commutation relation of
creation and annihilation operators is taken as equation (\ref{BEc})
and the statistical distribution is taken as the Bose-Einstein
distribution.

In this paper, we will construct an intermediate-statistics operator
realization for the angular momentum algebra, in which the maximum
occupation number is naturally embodied in the commutation relation
of creation and annihilation operators. Therefore, all results based
on this intermediate-statistics realization can naturally take the
influence of the restriction of occupation number into account. As a
comparison, the Holstein-Primakoff representation is a constrained
bosonic representation, i.e., a bosonic realization subject to a
constraint on the occupation number, so the influence of the maximum
occupation number cannot be taken into account directly.

From the intermediate-statistics realization for the angular
momentum algebra, we calculate the corresponding
intermediate-statistics distribution function. We show that the
statistical distribution that magnons obey is a modified
Bose-Einstein distribution.

Based on the intermediate-statistics realization for the angular
momentum algebra and the intermediate-statistics distribution
function, we calculate the dispersion relations of spin waves for
ferromagnet and antiferromagnet.

The magnons, as shown in the present paper, obey a kind of
intermediate statistics. As comparisons, we will discuss the
relations between the intermediate statistics that magnons obey and
the two important kinds of intermediate statistics: Haldane-Wu
fractional statistics \cite{gPauli} and the fractional statistics of
anyons \cite{Wilczek}. Haldane-Wu fractional statistics is
constructed based on the generalization of the Pauli exclusion
principle. The concept of anyons is introduced by analyzing the
symmetry properties of the wavefunction of identical particles: the
change of the phase factor of the wavefunction when two identical
particles exchange, instead of $+1$ or $-1$, is generalized to an
arbitrary phase factor $e^{i\theta}$. Each of these two kinds of
intermediate statistics has an intermediate-statistics parameter:
$g$ (Haldane-Wu) and $\alpha=\theta/\pi$ (anyon). The roles of $g$
and $\alpha$ in Haldane-Wu statistics and in the fractional
statistics of anyons are just as the role of the parameter $n$ in
the intermediate statistics obeyed by magnons. In this paper, we
will discuss the relations of $g$ and $n$ and $\alpha$ and $n$,
respectively.

A special case of the Heisenberg model, the one-dimensional $s=1/2$
Heisenberg
spin chain, can be solved exactly by the Bethe ansatz \cite{Bethe,Cini,Solyom}%
. By this exact solution, we can check the validity of our result
directly. Concretely, we will compare the exact spectrum obtained by
the Bethe ansatz method with the spectrum obtained by the
intermediate-statistics method and the spectrum obtained by the
Holstein-Primakoff method.

Moreover, we also compare our result with the experimental data of
$EuO$ given in \cite{GMP}. The result shows that at low temperatures
and low frequencies, the intermediate-statistics spin wave model is
more accurate than the bosonic spin wave model.

The present paper discusses the intermediate statistics and the
intermediate-statistics realization of angular momentum algebra and
their applications to magnetic systems. On the one hand, the
realization of angular momentum algebra has been of interest for a
long time. Besides the well-known Schwinger representation \cite{S}
and the Holstein-Primakoff representation \cite{HP}, there are many
other schemes, including the realization of $su\left(  2\right)  $
algebra \cite{OurPhysica,OurPRE,AKMS,Ruan,MSS}, and the realization
of $su_{q}\left(  2\right)  $ and $su_{q}\left(  n\right)  $ algebra
\cite{KK,SF}. On the other hand, the spin wave plays an important
role in magnetic problems \cite{Syr,WVW}. The concept of spin wave
has become a widely applied tool in the fields related to magnetism.
It has been applied to study magnetic semiconductors \cite{JKJK},
quasiequilibrium spin systems \cite{BD}, ballistic thermal transport
\cite{ZWL} and the thermodynamics \cite{Yam} in the Heisenberg spin
chain, one-dimensional ferromagnetic Bose gases \cite{ZCG},
spin-wave excitations in cylindrical ferromagnetic nanotubes
\cite{NC}, etc. In experiment, the properties of spin wave have been
directly measured \cite{GEFH,ZYLH}. The realizations of angular
momentum algebra are successful in describing magnetism in various
quantum systems \cite{TYV}. In the application of the realizations
of angular momentum algebra to magnetism, the realization is either
bosonic or fermionic. However, as shown above, magnons obey neither
Bose-Einstein nor Fermi-Dirac statistics. Therefore, a kind of
intermediate-statistics treatment is needed. As generalizations of
Bose-Einstein and Fermi-Dirac statistics, many schemes of
intermediate statistics have been discussed
\cite{gPauli,Wilczek,Bytsko,Gentile,OursAnn}.

In section \ref{operator}, we construct an intermediate-statistics
operator realization for the angular momentum algebra. In section
\ref{distributionfunction}, we calculate the intermediate-statistics
distribution function for magnons based on the commutation relation
between creation and annihilation operators of magnons. In sections
\ref{FSW} and \ref{aFWS}, we calculate the dispersion relations for
ferromagnetic and antiferromagnetic spin waves. In section
\ref{HWandanyon}, we discuss the relation between the intermediate
statistics that magnons obey and Haldane-Wu fractional statistics
and the relation between intermediate statistics and the fractional
statistics of anyons. In section \ref{Bethe}, we compare our result
with the exact solution obtained by the Bethe ansatz method. In
section \ref{numberical}, we compare the dispersion relation of a
ferromagnetic system, which is calculated based on intermediate
statistics, with the experimental data of $EuO$. The conclusions and
discussions are given in section \ref{conclusions}.

\section{Intermediate-statistics operator realization for angular momentum
algebra\label{operator}}

For solving the spectrum of a magnetic system with Hamiltonian
(\ref{Hmg}), we need a representation of angular momentum algebra.
As discussed above, the Holstein-Primakoff representation is a
bosonic realization with a putting-in-by-hand restriction on the
occupation number, and when using it to solve the spectrum, since
only the operator relation is taken into account, the information of
the restriction on the occupation number is ignored. For taking the
influence of the restriction on the occupation number into account,
we need a representation in which the information of the maximum
occupation number is embodied in the operator relation rather than
put in an additional restriction by hand. In this section, we
construct an intermediate-statistics operator realization for the
angular momentum algebra, in which the maximum occupation number is
naturally embodied in the commutation relation of creation and
annihilation operators.

In the case of the spin wave, we only focus on the low-lying
excitation. In other words, in our case the spin deviation
$N_{\mathbf{\ell}}$ is always very small, and then equation
(\ref{spin}) can be expanded around $N_{\mathbf{\ell }}=0$; only
taking the next-to-leading-order contribution into account, we
have%
\begin{align}
S_{\mathbf{\ell}}^{+}\left\vert N_{\mathbf{\ell}}\right\rangle  &
=\sqrt
{2S}\left(  1-\frac{N_{\mathbf{\ell}}-1}{4S}\right)  \sqrt{N_{\mathbf{\ell}}%
}\left\vert N_{\mathbf{\ell}}-1\right\rangle ,\nonumber\\
S_{\mathbf{\ell}}^{-}\left\vert N_{\mathbf{\ell}}\right\rangle  &
=\sqrt {2S}\sqrt{N_{\mathbf{\ell}}+1}\left(
1-\frac{N_{\mathbf{\ell}}}{4S}\right)
\left\vert N_{\mathbf{\ell}}+1\right\rangle ,\nonumber\\
S_{\mathbf{\ell}}^{z}\left\vert N_{\mathbf{\ell}}\right\rangle  &
=S-N_{\mathbf{\ell}}.
\end{align}

For constructing a realization of angular momentum algebra, we first
introduce
an intermediate-statistics operator realization:%
\begin{align}
\left[
a_{\mathbf{\ell}},a_{\mathbf{\ell}^{\prime}}^{\dagger}\right]   &
=\left\{  \frac{1-\frac{N_{\mathbf{\ell}}}{n}}{\left(
1-\frac{N_{\mathbf{\ell
}}}{2n}\right)  ^{2}}+N_{\mathbf{\ell}}\left[  \frac{1-\frac{N_{\mathbf{\ell}%
}}{n}}{\left(  1-\frac{N_{\mathbf{\ell}}}{2n}\right)  ^{2}}-\frac
{1-\frac{N_{\mathbf{\ell}}-1}{n}}{\left(  1-\frac{N_{\mathbf{\ell}}-1}%
{2n}\right)  ^{2}}\right]  \right\}  \delta_{\mathbf{\ell\ell}^{\prime}%
},\nonumber\\
\left[
N_{\mathbf{\ell}},a_{\mathbf{\ell}^{\prime}}^{\dagger}\right]   &
=a_{\mathbf{\ell}}^{\dagger}\delta_{\mathbf{\ell\ell}^{\prime}}\text{\
and }\left[  N_{\mathbf{\ell}},a_{\mathbf{\ell}^{\prime}}\right]
=-a_{\mathbf{\ell}}\delta_{\mathbf{\ell\ell}^{\prime}},\label{icr}%
\end{align}
where $n=2S$. It can be directly checked that such an operator
realization corresponds to a kind of intermediate statistics with a
maximum occupation number $n$.

By creation, annihilation, and number operators,
$a_{\mathbf{\ell}}$, $a_{\mathbf{\ell}}^{\dagger}$, and
$N_{\mathbf{\ell}}$, we can construct an
intermediate-statistics realization of angular momentum algebra:%
\begin{align}
S_{\mathbf{\ell}}^{+}  &  =\sqrt{2S}\left(  1-\frac{N_{\mathbf{\ell}}}%
{4S}\right)  a_{\mathbf{\ell}},\nonumber\\
S_{\mathbf{\ell}}^{-}  &
=\sqrt{2S}a_{\mathbf{\ell}}^{\dagger}\left(
1-\frac{N_{\mathbf{\ell}}}{4S}\right)  ,\nonumber\\
S_{\mathbf{\ell}}^{z}  &  =S-N_{\mathbf{\ell}}.\label{amreal}%
\end{align}
It can be directly checked that, with the commutation relation
(\ref{icr}), $S_{\mathbf{\ell}}^{+}$, $S_{\mathbf{\ell}}^{-}$, and
$S_{\mathbf{\ell}}^{z}$ satisfy the operator relation of angular
momentum:
\begin{align}
\left[
S_{\mathbf{\ell}}^{z},S_{\mathbf{\ell}^{\prime}}^{\pm}\right]   &
=\pm S_{\mathbf{\ell}}^{\pm}\delta_{\mathbf{\ell\ell}^{\prime}},\nonumber\\
\left[  S_{\mathbf{\ell}}^{+},S_{\mathbf{\ell}^{\prime}}^{-}\right]
&
=2S_{\mathbf{\ell}}^{z}\delta_{\mathbf{\ell\ell}^{\prime}}.\label{am}%
\end{align}
In this scheme, $a_{\mathbf{\ell}}^{\dagger}$ and
$a_{\mathbf{\ell}}$ are the creation and annihilation operators of a
localized spin deviation at a single site, and $N_{\mathbf{\ell}}$
is the spin deviation operator which can be
expressed as%
\begin{equation}
N_{\mathbf{\ell}}=\frac{a_{\mathbf{\ell}}^{\dagger}a_{\mathbf{\ell}}+2n\left[
a_{\mathbf{\ell}}^{\dagger}a_{\mathbf{\ell}}+\left(  n+1\right)
-\sqrt{\left(  n+1\right)  ^{2}-\left(  2n+1\right)  a_{\mathbf{\ell}%
}^{\dagger}a_{\mathbf{\ell}}}\right]  }{a_{\mathbf{\ell}}^{\dagger
}a_{\mathbf{\ell}}+4n}.\label{Nl}%
\end{equation}
It should be emphasized that in such intermediate statistics,
$N_{\mathbf{\ell }}\neq
a_{\mathbf{\ell}}^{\dagger}a_{\mathbf{\ell}}$. Concretely, for a
given set of creation and annihilation operators,
$a_{\mathbf{\ell}}^{\dagger}$ and $a_{\mathbf{\ell}}$, the
corresponding number operator $N_{\mathbf{\ell}}$ can be constructed
from the operator relations of $a_{\mathbf{\ell}}^{\dagger}$,
$a_{\mathbf{\ell}}^{\dagger}$, and $N_{\mathbf{\ell}}$. As shown in
\cite{OurPhysica} and \cite{OurPRE}, the number operator of
intermediate
statistics in general cannot take the form of $N_{\mathbf{\ell}}%
=a_{\mathbf{\ell}}^{\dagger}a_{\mathbf{\ell}}$. The realization of
the number operator (\ref{Nl}) shows that the spin deviation
corresponds to a kind of intermediate statistics.

For describing the nonlocalized excitations of such a magnetic
system, taking translational symmetry into account, we replace the
creation (annihilation) operator $a_{\mathbf{\ell}}^{\dagger}$
($a_{\mathbf{\ell}}$) which creates (annihilates) localized spin
deviations with the creation (annihilation) operator
$b_{\mathbf{k}}^{\dagger}$ ($b_{\mathbf{k}}$) which creates
(annihilates) nonlocalized excitations by the transformation%
\begin{align}
a_{\mathbf{\ell}}  &  =\frac{1}{\sqrt{W}}\sum_{\mathbf{k}}e^{i\mathbf{k}%
\cdot\mathbf{\ell}}b_{\mathbf{k}},\nonumber\\
a_{\mathbf{\ell}}^{\dagger}  &  =\frac{1}{\sqrt{W}}\sum_{\mathbf{k}%
}e^{-i\mathbf{k}\cdot\mathbf{\ell}}b_{\mathbf{k}}^{\dagger},\label{atob}%
\end{align}
where $W$ is the number of lattice sites. Then%
\begin{align}
\left[  b_{\mathbf{k}},b_{\mathbf{k}^{\prime}}^{\dagger}\right]   &
=\left\{ \frac{1-\frac{N_{\mathbf{k}}}{n}}{\left(
1-\frac{N_{\mathbf{k}}}{2n}\right) ^{2}}+N_{\mathbf{k}}\left[
\frac{1-\frac{N_{\mathbf{k}}}{n}}{\left(
1-\frac{N_{\mathbf{k}}}{2n}\right)  ^{2}}-\frac{1-\frac{N_{\mathbf{k}}-1}{n}%
}{\left(  1-\frac{N_{\mathbf{k}}-1}{2n}\right)  ^{2}}\right]
\right\}
\delta_{\mathbf{kk}^{\prime}},\nonumber\\
\left[  N_{\mathbf{k}},b_{\mathbf{k}^{\prime}}^{\dagger}\right]   &
=b_{\mathbf{k}}^{\dagger}\delta_{\mathbf{kk}^{\prime}}\text{\ and
}\left[ N_{\mathbf{k}},b_{\mathbf{k}^{\prime}}\right]
=-b_{\mathbf{k}}\delta
_{\mathbf{kk}^{\prime}}.\label{bdbformagnon}%
\end{align}
$N_{\mathbf{k}}$, $b_{\mathbf{k}}^{\dagger}$, and $b_{\mathbf{k}}$
are the number, creation, and annihilation operators of magnons,
respectively, which describe the nonlocalized elementary excitation
of the system.

The above result shows that magnons, the elementary excitation of a
magnetic system, obey the intermediate statistics defined by
(\ref{bdbformagnon}), which has a maximum occupation number $n$.
Only when $n\rightarrow\infty$, such intermediate statistics returns
to Bose-Einstein statistics. In other words, the spin wave is
essentially an intermediate-statistics type elementary excitation.

\section{The intermediate-statistics distribution
function\label{distributionfunction}}

When calculating the spectrum of a magnetic system, one needs to use
the statistical distribution function of the magnon. In the common
treatment, the statistical distribution is approximately taken as
the Bose-Einstein distribution. However, as discussed above, the
magnon in fact obeys intermediate statistics. In this section, we
seek for the statistical distribution for an ideal magnon gas based
on the commutation relation (\ref{bdbformagnon}).

The average particle number of $\mathbf{k}$ state can be obtained by%

\begin{equation}
\left\langle N_{\mathbf{k}}\right\rangle =\frac{1}{Z}Tr\left[
e^{-\beta
\left(  H-\mu N\right)  }N_{\mathbf{k}}\right]  ,\label{Tr}%
\end{equation}
where $Z=Tre^{-\beta\left(  H-\mu N\right)  }$, $N$ is the total
particle number operator, and $\mu$ is the chemical potential. For
the low-lying excitation, we can expand the expression of the
particle number operator,
similar to equation (\ref{Nl}), as%
\begin{equation}
N_{\mathbf{k}}\simeq\frac{\left(  2n+1\right)  ^{2}}{4n\left(
n+1\right)
}b_{\mathbf{k}}^{\dagger}b_{\mathbf{k}}-\frac{\left(  2n+1\right)  ^{3}%
}{16n^{2}\left(  n+1\right)  ^{3}}\left(  b_{\mathbf{k}}^{\dagger
}b_{\mathbf{k}}\right)  ^{2}.\label{Nl-2}%
\end{equation}
By the relation%
\begin{equation}
e^{-\beta\left(  H-\mu N\right)
}b_{\mathbf{k}}^{\dagger}=e^{-\beta\left(
\varepsilon_{\mathbf{k}}-\mu\right)
}b_{\mathbf{k}}^{\dagger}e^{-\beta\left( H-\mu N\right)  },
\end{equation}
from (\ref{Tr}), we achieve%
\begin{align}
\left\langle N_{\mathbf{k}}\right\rangle  &  =e^{-\beta\left(
\varepsilon
_{\mathbf{k}}-\mu\right)  }\frac{1}{Z}\frac{\left(  2n+1\right)  ^{2}%
}{4n\left(  n+1\right)  }\nonumber\\
\times &  \left\{  Tr\left[  e^{-\beta\left(  H-\mu N\right)  }b_{\mathbf{k}%
}b_{\mathbf{k}}^{\dagger}\right]  -\frac{2n+1}{4n\left(  n+1\right)  ^{2}%
}Tr\left[  e^{-\beta\left(  H-\mu N\right)  }b_{\mathbf{k}}b_{\mathbf{k}%
}^{\dagger}b_{\mathbf{k}}b_{\mathbf{k}}^{\dagger}\right]  \right\}
,
\end{align}
where $\varepsilon_{\mathbf{k}}$ is the energy of $\mathbf{k}$
state. Based on
the operator relation obtained in the above section, we construct%
\begin{equation}
b_{\mathbf{k}}^{\dagger}b_{\mathbf{k}}=N_{\mathbf{k}}\frac{1-\frac
{N_{\mathbf{k}}-1}{n}}{\left(  1-\frac{N_{\mathbf{k}}-1}{2n}\right)  ^{2}%
}\text{ \ \ and \ \ }b_{\mathbf{k}}b_{\mathbf{k}}^{\dagger}=\left(
N_{\mathbf{k}}+1\right)  \frac{1-\frac{N_{\mathbf{k}}}{n}}{\left(
1-\frac{N_{\mathbf{k}}}{2n}\right)  ^{2}}.\label{bbN}%
\end{equation}
Ignoring $\left\langle N_{\mathbf{k}}^{2}\right\rangle $, we can
solve
$\left\langle N_{\mathbf{k}}\right\rangle $:%
\begin{equation}
\left\langle N_{\mathbf{k}}\right\rangle =\displaystyle\frac{1-\frac
{4n^{3}+4n^{2}+2n+1}{16n^{2}\left(  n+1\right)
^{3}}}{e^{\beta\left(
\varepsilon_{\mathbf{k}}-\mu\right)  }-\left[  1-\frac{6n^{3}+8n^{2}%
+4n+1}{8n^{2}\left(  n+1\right)  ^{3}}\right]  }.\label{distribution}%
\end{equation}
It can be directly seen from this equation that the statistical
distribution defined by the commutation relation
(\ref{bdbformagnon}) is a modified Bose-Einstein distribution, and
when $n\rightarrow\infty$, equation (\ref{distribution}) returns to
the Bose-Einstein distribution. In other words, the spin wave obeys
a modified Bose-Einstein distribution.

\section{Intermediate-statistics ferromagnetic spin waves\label{FSW}}

In this section, we calculate the dispersion relation of a
ferromagnetic spin
wave. The Hamiltonian of the Heisenberg model reads%
\begin{equation}
H=-J_{1}\sum_{\mathbf{\ell},\mathbf{\delta}_{1}}\mathbf{S}_{\mathbf{\ell}%
}\cdot\mathbf{S}_{\mathbf{\ell}+\mathbf{\delta}_{1}}-J_{2}\sum_{\mathbf{\ell
},\mathbf{\delta}_{2}}\mathbf{S}_{\mathbf{\ell}}\cdot\mathbf{S}_{\mathbf{\ell
}+\mathbf{\delta}_{2}}+\cdots,\label{Hn-nn}%
\end{equation}
where $\mathbf{\delta}_{1}$ and $\mathbf{\delta}_{2}$ connect spin
$\mathbf{\ell}$ with its nearest and next-to-nearest neighbors and
$J_{1}$ and $J_{2}$ denote the exchange parameters corresponding to
the nearest-neighbor and next-to-nearest-neighbor couplings.

We first consider the nearest-neighbor contribution. The Hamiltonian reads%
\begin{equation}
H=-J_{1}\sum_{\mathbf{\ell},\mathbf{\delta}_{1}}\left[  S_{\mathbf{\ell}}%
^{z}S_{\mathbf{\ell}+\mathbf{\delta}_{1}}^{z}+\frac{1}{2}\left(
S_{\mathbf{\ell}}^{+}S_{\mathbf{\ell}+\mathbf{\delta}_{1}}^{-}+S_{\mathbf{\ell
}}^{-}S_{\mathbf{\ell}+\mathbf{\delta}_{1}}^{+}\right)  \right]  .\label{Hnn}%
\end{equation}
Substituting the intermediate-statistics representation of angular
momentum
algebra (\ref{amreal}) into (\ref{Hnn}), we have%
\begin{equation}
H=H_{0}+H_{2}+H_{4}%
\end{equation}
with%
\begin{equation}
H_{0}=-J_{1}\frac{n^{2}}{4}\sum_{\mathbf{\ell},\mathbf{\delta}_{1}}%
1=-J_{1}WZ_{1}\frac{n^{2}}{4},
\end{equation}

\begin{equation}
H_{2}=J_{1}\sum_{\mathbf{\ell},\mathbf{\delta}_{1}}\left[  nN_{\mathbf{\ell}%
}-\frac{n}{2}\left(  a_{\mathbf{\ell}}a_{\mathbf{\ell}+\mathbf{\delta}_{1}%
}^{\dagger}+a_{\mathbf{\ell}}^{\dagger}a_{\mathbf{\ell}+\mathbf{\delta}_{1}%
}\right)  \right]  ,\label{H2}%
\end{equation}
and%
\begin{equation}
H_{4}=-J_{1}\sum_{\mathbf{\ell},\mathbf{\delta}_{1}}\left[  N_{\mathbf{\ell}%
}N_{\mathbf{\ell}+\mathbf{\delta}_{1}}-\frac{1}{4}\left(  a_{\mathbf{\ell}%
}a_{\mathbf{\ell}+\mathbf{\delta}_{1}}^{\dagger}N_{\mathbf{\ell}%
+\mathbf{\delta}_{1}}+N_{\mathbf{\ell}}a_{\mathbf{\ell}}a_{\mathbf{\ell
}+\mathbf{\delta}_{1}}^{\dagger}+a_{\mathbf{\ell}}^{\dagger}N_{\mathbf{\ell
}+\mathbf{\delta}_{1}}a_{\mathbf{\ell}+\mathbf{\delta}_{1}}+a_{\mathbf{\ell}%
}^{\dagger}N_{\mathbf{\ell}}a_{\mathbf{\ell}+\mathbf{\delta}_{1}}\right)
\right]  ,\label{H4}%
\end{equation}
where $Z_{1}$ is the number of nearest neighbors.

In the following, by the above operator relations and the
statistical distribution function given in sections \ref{operator}
and \ref{distributionfunction}, we can calculate the spectrum
directly.

\subsection{The quadratic contribution}

First, we calculate the contribution from the terms quadratic in the
creation
and annihilation operators. Substituting (\ref{atob}) into (\ref{H2}) gives%
\begin{equation}
H_{2}=J_{1}Z_{1}n\sum_{\mathbf{k}}N_{\mathbf{k}}-J_{1}Z_{1}\frac{n}{2}%
\sum_{\mathbf{k}}\gamma_{\mathbf{k}}^{N}\left(
b_{\mathbf{k}}^{\dagger
}b_{\mathbf{k}}+b_{\mathbf{k}}b_{\mathbf{k}}^{\dagger}\right)  .\label{H2b}%
\end{equation}
Here $\gamma_{\mathbf{k}}^{N}$ is defined to be $\gamma_{\mathbf{k}}^{N}%
=\frac{1}{Z_{1}}\sum_{\mathbf{\delta}_{1}}e^{i\mathbf{k}\cdot\mathbf{\delta
}_{1}}$ with $\gamma_{\mathbf{k}}^{N}=\gamma_{-\mathbf{k}}^{N}$ due
to the symmetry. Then, substituting (\ref{bbN}) into (\ref{H2b}) and
preserving only
the first-order contribution of $N_{\mathbf{k}}$ gives%
\begin{equation}
H_{2}=J_{1}Z_{1}n\sum_{\mathbf{k}}\left[  \left(  1-\gamma_{\mathbf{k}}%
^{N}\right)  +\frac{1}{2\left(  2n+1\right)  ^{2}}\gamma_{\mathbf{k}}%
^{N}\right]  N_{\mathbf{k}}.
\end{equation}
\newline We achieve the dispersion relation of magnons to the second order:
\begin{equation}
\hbar\omega_{\mathbf{k}}^{\left(  2\right)  }=J_{1}Z_{1}n\left[
\left(
1-\gamma_{\mathbf{k}}^{N}\right)  +\frac{1}{2\left(  2n+1\right)  ^{2}}%
\gamma_{\mathbf{k}}^{N}\right]  .\label{spectrum2}%
\end{equation}
The second term in (\ref{spectrum2}) is the modification coming from
the influence of the restriction on the occupation number.

\subsection{The quartic contribution: the interaction between magnons}

The contribution from the terms quartic in the creation and
annihilation operators describes the interaction between magnons. A
similar treatment can also be used to deal with the quartic terms.

From equation (\ref{H4}), by the operator relations given above, up
to quartic
terms, we obtain%
\begin{align}
H_{4}  &  =-J_{1}\frac{Z_{1}}{W}\sum_{\mathbf{k}_{1}\mathbf{k}_{2}%
\mathbf{k}_{3}\mathbf{k}_{4}}\left\{  \left[  1+\frac{1}{4n\left(
n+1\right)
}\right]  ^{2}\delta_{\mathbf{k}_{1}-\mathbf{k}_{2}+\mathbf{k}_{3}%
,\mathbf{k}_{4}}\gamma_{\mathbf{k}_{3}-\mathbf{k}_{4}}^{N}b_{\mathbf{k}_{1}%
}^{\dagger}b_{\mathbf{k}_{2}}b_{\mathbf{k}_{3}}^{\dagger}b_{\mathbf{k}_{4}%
}\right. \nonumber\\
&  -\frac{1}{4}\left[  1+\frac{1}{4n\left(  n+1\right)  }\right]
\left(
\delta_{\mathbf{k}_{1}-\mathbf{k}_{2}-\mathbf{k}_{3},-\mathbf{k}_{4}}%
\gamma_{\mathbf{k}_{1}}^{N}b_{\mathbf{k}_{1}}b_{\mathbf{k}_{2}}^{\dagger
}b_{\mathbf{k}_{3}}^{\dagger}b_{\mathbf{k}_{4}}+\delta_{\mathbf{k}%
_{1}-\mathbf{k}_{2}-\mathbf{k}_{3},-\mathbf{k}_{4}}\gamma_{\mathbf{k}_{4}}%
^{N}b_{\mathbf{k}_{1}}^{\dagger}b_{\mathbf{k}_{2}}b_{\mathbf{k}_{3}%
}b_{\mathbf{k}_{4}}^{\dagger}\right. \nonumber\\
&  \left.  \left.  +\delta_{\mathbf{k}_{1}+\mathbf{k}_{2}-\mathbf{k}%
_{3},\mathbf{k}_{4}}\gamma_{\mathbf{k}_{1}}^{N}b_{\mathbf{k}_{1}}^{\dagger
}b_{\mathbf{k}_{2}}^{\dagger}b_{\mathbf{k}_{3}}b_{\mathbf{k}_{4}}%
+\delta_{\mathbf{k}_{1}+\mathbf{k}_{2}-\mathbf{k}_{3},\mathbf{k}_{4}}%
\gamma_{\mathbf{k}_{4}}^{N}b_{\mathbf{k}_{1}}^{\dagger}b_{\mathbf{k}_{2}%
}^{\dagger}b_{\mathbf{k}_{3}}b_{\mathbf{k}_{4}}\right)  \right\}  .
\end{align}
For long-wavelength spin waves, the main contribution comes from the
interactions that do not change the state of the spin wave. For
example, for
the term in proportion to $b_{\mathbf{k}_{1}}^{\dagger}b_{\mathbf{k}_{2}%
}^{\dagger}b_{\mathbf{k}_{3}}b_{\mathbf{k}_{4}}$, only the
contributions
corresponding to $\mathbf{k}_{1}=\mathbf{k}_{3}$ and $\mathbf{k}%
_{2}=\mathbf{k}_{4}$, or $\mathbf{k}_{1}=\mathbf{k}_{4}$ and $\mathbf{k}%
_{2}=\mathbf{k}_{3}$ remain \cite{Solyom}. Only taking these
contributions
into account, we approximately achieve%
\begin{align}
H_{4}  &  \simeq-J_{1}\frac{Z_{1}}{W}\sum_{\mathbf{k}_{1}\mathbf{k}_{2}%
}\left\{  \left[  1+\frac{1}{4n\left(  n+1\right)  }\right]
^{2}\left(
\gamma_{0}^{N}b_{\mathbf{k}_{1}}^{\dagger}b_{\mathbf{k}_{1}}b_{\mathbf{k}_{2}%
}^{\dagger}b_{\mathbf{k}_{2}}+\gamma_{\mathbf{k}_{2}-\mathbf{k}_{1}}%
^{N}b_{\mathbf{k}_{1}}^{\dagger}b_{\mathbf{k}_{1}}b_{\mathbf{k}_{2}%
}b_{\mathbf{k}_{2}}^{\dagger}\right)  \right. \nonumber\\
&  \left.  -\frac{1}{4}\left[  1+\frac{1}{4n\left(  n+1\right)
}\right]
\left[  \gamma_{\mathbf{k}_{1}}^{N}b_{\mathbf{k}_{1}}b_{\mathbf{k}_{1}%
}^{\dagger}b_{\mathbf{k}_{2}}^{\dagger}b_{\mathbf{k}_{2}}+3\gamma
_{\mathbf{k}_{2}}^{N}b_{\mathbf{k}_{1}}^{\dagger}b_{\mathbf{k}_{1}%
}b_{\mathbf{k}_{2}}b_{\mathbf{k}_{2}}^{\dagger}+\left(  3\gamma_{\mathbf{k}%
_{1}}^{N}+\gamma_{\mathbf{k}_{2}}^{N}\right)
b_{\mathbf{k}_{1}}^{\dagger
}b_{\mathbf{k}_{1}}b_{\mathbf{k}_{2}}^{\dagger}b_{\mathbf{k}_{2}}\right]
\right\}  .
\end{align}

By equation (\ref{bbN}), up to $N_{\mathbf{k}}^{2}$, we obtain%
\begin{align}
H  &
=-J_{1}\frac{Z_{1}}{W}\sum_{\mathbf{k}_{1}\mathbf{k}_{2}}\left[
\left(
1-\gamma_{\mathbf{k}_{1}}^{N}\right)  \left(  1-\gamma_{\mathbf{k}_{2}}%
^{N}\right)  \right. \\
&  +\left.  \frac{\gamma_{\mathbf{k}_{1}}^{N}+\gamma_{\mathbf{k}_{2}}%
^{N}+2\gamma_{\mathbf{k}_{1}}^{N}\gamma_{\mathbf{k}_{2}}^{N}}{2\left(
2n+1\right)  ^{2}}+\frac{\gamma_{\mathbf{k}_{1}}^{N}\gamma_{\mathbf{k}_{2}%
}^{N}}{4n\left(  n+1\right)  \left(  2n+1\right)  ^{2}}\right]  N_{\mathbf{k}%
_{1}}N_{\mathbf{k}_{2}}.
\end{align}
In the calculation, the relations $\gamma_{0}^{N}=1$, $\gamma_{\mathbf{k}%
_{2}-\mathbf{k}_{1}}^{N}=\gamma_{\mathbf{k}_{2}}^{N}\gamma_{\mathbf{k}_{1}%
}^{N}$ \cite{Majlis}, $\sum_{\mathbf{k}}\gamma_{\mathbf{k}}^{N}=0$,
and
$\sum_{\mathbf{k}_{1}\mathbf{k}_{2}}\gamma_{\mathbf{k}_{1}}^{N}N_{\mathbf{k}%
_{1}}N_{\mathbf{k}_{2}}=\sum_{\mathbf{k}_{1}\mathbf{k}_{2}}\gamma
_{\mathbf{k}_{2}}^{N}N_{\mathbf{k}_{1}}N_{\mathbf{k}_{2}}$\textbf{\
}have been used.

When the number of excited magnons fluctuates little, we can take
the
approximation \cite{Solyom}%
\begin{equation}
N_{\mathbf{k}_{1}}N_{\mathbf{k}_{2}}\simeq\left\langle N_{\mathbf{k}_{1}%
}\right\rangle N_{\mathbf{k}_{2}}+N_{\mathbf{k}_{1}}\left\langle
N_{\mathbf{k}_{2}}\right\rangle -\left\langle
N_{\mathbf{k}_{1}}\right\rangle \left\langle
N_{\mathbf{k}_{2}}\right\rangle .
\end{equation}
Then%
\begin{align}
H_{4}  &  \simeq-2J_{1}\frac{Z_{1}}{W}\sum_{\mathbf{qk}}\left[
\left( 1-\gamma_{\mathbf{q}}^{N}\right)  \left(
1-\gamma_{\mathbf{k}}^{N}\right)
+\frac{\gamma_{\mathbf{q}}^{N}+\gamma_{\mathbf{k}}^{N}+2\gamma_{\mathbf{q}%
}^{N}\gamma_{\mathbf{k}}^{N}}{2\left(  2n+1\right)  ^{2}}\right. \nonumber\\
&  +\left.
\frac{\gamma_{\mathbf{q}}^{N}\gamma_{\mathbf{k}}^{N}}{4n\left(
n+1\right)  \left(  2n+1\right)  ^{2}}\right]  \left\langle N_{\mathbf{q}%
}\right\rangle N_{\mathbf{k}}+J_{1}\frac{Z_{1}}{W}\sum_{\mathbf{k}%
_{1}\mathbf{k}_{2}}\left[  \left(
1-\gamma_{\mathbf{k}_{1}}^{N}\right)
\left(  1-\gamma_{\mathbf{k}_{2}}^{N}\right)  \right. \nonumber\\
&  +\left.  \frac{\gamma_{\mathbf{k}_{1}}^{N}+\gamma_{\mathbf{k}_{2}}%
^{N}+2\gamma_{\mathbf{k}_{1}}^{N}\gamma_{\mathbf{k}_{2}}^{N}}{2\left(
2n+1\right)  ^{2}}+\frac{\gamma_{\mathbf{k}_{1}}^{N}\gamma_{\mathbf{k}_{2}%
}^{N}}{4n\left(  n+1\right)  \left(  2n+1\right)  ^{2}}\right]
\left\langle N_{\mathbf{k}_{1}}\right\rangle \left\langle
N_{\mathbf{k}_{2}}\right\rangle .
\end{align}
Consequently, the contribution from the interaction between magnons
to the dispersion relation reads
\begin{align}
\hbar\omega_{\mathbf{k}}^{\left(  4\right)  }  &  =-2J_{1}\frac{Z_{1}}%
{W}\left\{  \sum_{\mathbf{q}}\left[  \left(
1-\gamma_{\mathbf{q}}^{N}\right)
+\frac{\gamma_{\mathbf{q}}^{N}}{2\left(  2n+1\right)  ^{2}}\right]
\left\langle N_{\mathbf{q}}\right\rangle \right. \nonumber\\
&  \left.  -\gamma_{\mathbf{k}}^{N}\sum_{\mathbf{q}}\left[  \left(
1-\gamma_{\mathbf{q}}^{N}\right)  -\frac{1}{2\left(  2n+1\right)  ^{2}}%
-\frac{\gamma_{\mathbf{q}}^{N}}{4n\left(  n+1\right)  }\right]
\left\langle
N_{\mathbf{q}}\right\rangle \right\}  .\label{spectrum4}%
\end{align}
It should be emphasized that the statistical distribution function
$\left\langle N_{\mathbf{q}}\right\rangle $ is the
intermediate-statistical distribution given by (\ref{distribution}),
rather than the Bose-Einstein distribution as that in the common
treatment.

\subsection{Comparison with the result of the Holstein-Primakoff
representation}

From the above discussion on the commutation relation of creation
and annihilation operators, we have already known that magnons obey
a kind of intermediate statistics with a maximum occupation number
$n$, and when $n\rightarrow\infty$ this intermediate statistics
returns to Bose-Einstein statistics. In the common treatment, one
approximately assumes that magnons obey Bose-Einstein statistics.
Under this approximation, the dispersion
relation reads \cite{Solyom}%
\begin{equation}
\hbar\omega_{\mathbf{k}}=J_{1}Z_{1}2S\left(
1-\gamma_{\mathbf{k}}^{N}\right) -2J_{1}\frac{Z_{1}}{W}\left(
1-\gamma_{\mathbf{k}}^{N}\right)  \left[
\sum_{\mathbf{q}}\left\langle N_{\mathbf{q}}^{Bose}\right\rangle
-\sum_{\mathbf{q}}\gamma_{\mathbf{q}}^{N}\left\langle N_{\mathbf{q}}%
^{Bose}\right\rangle \right]  ,\label{HPresult}%
\end{equation}
where $\left\langle N_{\mathbf{q}}^{Bose}\right\rangle $ denotes the
Bose-Einstein distribution.

Comparing the dispersion relation (\ref{HPresult}) with the
dispersion relation (\ref{spectrum2}) and (\ref{spectrum4}), we can
see that when replacing Bose-Einstein statistics by intermediate
statistics, some additional modifications relying on the maximum
occupation number $n$ appear. When the maximum occupation number
$n\rightarrow\infty$, such modifications vanish.

Especially, the statistical distribution function $\left\langle N_{\mathbf{q}%
}^{Bose}\right\rangle $ in the Holstein-Primakoff result
(\ref{HPresult}) is the Bose-Einstein distribution, while in the
present result (\ref{spectrum4}),
the statistical distribution function $\left\langle N_{\mathbf{q}%
}\right\rangle $ is the intermediate-statistics distribution
(\ref{distribution}) which is a modified Bose-Einstein distribution.

\subsection{Other contributions}

For comparison with the experiment, we need to consider all effects
as possible. In this subsection, we discuss the contribution from
next-to-nearest neighbors and dipolar interactions.

\subsubsection{The next-to-nearest-neighbor contribution}

The contribution from the next-to-nearest-neighbor coupling can be
calculated
directly by the same procedure:%
\begin{align}
\hbar\omega_{\mathbf{k}}^{NN}  &  =J_{2}Z_{2}n\left[  \left(
1-\gamma
_{\mathbf{k}}^{NN}\right)  +\frac{1}{2\left(  2n+1\right)  ^{2}}%
\gamma_{\mathbf{k}}^{NN}\right] \nonumber\\
&  -2J_{2}\frac{Z_{2}}{W}\left\{  \sum_{\mathbf{q}}\left[  \left(
1-\gamma_{\mathbf{q}}^{NN}\right)
+\frac{\gamma_{\mathbf{q}}^{NN}}{2\left( 2n+1\right)  ^{2}}\right]
\left\langle N_{\mathbf{q}}\right\rangle \right.
\nonumber\\
&  -\left.  \gamma_{\mathbf{k}}^{NN}\sum_{\mathbf{q}}\left[  \left(
1-\gamma_{\mathbf{q}}^{NN}\right)  -\frac{1}{2\left(  2n+1\right)  ^{2}}%
-\frac{\gamma_{\mathbf{q}}^{NN}}{4n\left(  n+1\right)  }\right]
\left\langle N_{\mathbf{q}}\right\rangle \right\}  .
\end{align}
where $Z_{2}$ is the number of next-to-nearest neighbors and $\gamma
_{\mathbf{k}}^{NN}=\frac{1}{Z_{2}}\sum_{\mathbf{\delta}_{2}}e^{i\mathbf{k}%
\cdot\mathbf{\delta}_{2}}$.

\subsubsection{The dipolar interaction}

In the above, we only take the contribution from the exchange
coupling into account. Besides the exchange coupling, there still
exists a classical dipolar interaction which is caused by the
interaction between the magnetic moments.
The Hamiltonian of the dipolar interaction reads%
\begin{equation}
H_{dip}=\frac{1}{2}g^{2}\mu_{B}^{2}\sum_{i,j}\left[  \frac{\mathbf{S}_{i}%
\cdot\mathbf{S}_{j}}{r_{ij}^{3}}-\frac{3\left(  \mathbf{S}_{i}\cdot
\mathbf{r}_{ij}\right)  \left(
\mathbf{S}_{j}\cdot\mathbf{r}_{ij}\right)
}{r_{ij}^{5}}\right]  ,\label{Hdip}%
\end{equation}
where $\mu_{B}$ is the Bohr magnon and $g=2$ is the Land\'{e}
factor. In principle, we need to substitute the
intermediate-statistics operator realization of angular momentum
algebra (\ref{amreal}) and (\ref{icr}) into this Hamiltonian, and,
then, calculate the influence of the intermediate-statistics dipolar
interaction to the dispersion relation. However, since the
contribution from the dipolar interaction, in comparison with the
contribution from the exchange coupling, is small, we ignore the
intermediate-statistics modification to the dipolar contribution,
i.e., for the dipolar interaction, we approximately use the result
obtained by the Holstein-Primakoff representation.

Under the assumption of isotropy, the contribution from the dipolar
interaction, based on the result given in \cite{White}, can be
approximately
expressed as%
\begin{equation}
\hbar\omega_{\mathbf{k}}^{dip}=\frac{4\pi}{3}g\mu_{B}M\left(
T\right)  ,
\end{equation}
where $M\left(  T\right)  $ is the magnetization. Then the
dispersion relation reads
\begin{equation}
\hbar\omega_{\mathbf{k}}^{total}=\hbar\omega_{\mathbf{k}}+\hbar\omega
_{\mathbf{k}}^{dip},\label{all}%
\end{equation}
where $\hbar\omega_{\mathbf{k}}$ comes from the exchange coupling
and $\hbar\omega_{\mathbf{k}}^{dip}$ comes from the dipolar
interaction.

\section{Intermediate-statistics antiferromagnetic spin waves \label{aFWS}}

It has been shown that there do exist quantized spin waves in
antiferromagnets \cite{WVW}. In this section, we calculate the
dispersion relation for antiferromagnetic spin waves based on the
intermediate-statistics scheme.

For antiferromagnets, the spin structure of the crystal is
considered as two interpenetrating sublattices $A$ and $B$ with the
property that all nearest neighbors of a spin on $A$ lie on $B$, and
\textit{vice versa}. The
Hamiltonian reads%
\begin{equation}
H=2J\sum_{\mathbf{\ell}\mathbf{\delta}}\left[  S_{A\mathbf{\ell}}%
^{z}S_{B\mathbf{\ell}+\mathbf{\delta}}^{z}+\frac{1}{2}\left(
S_{A\mathbf{\ell
}}^{+}S_{B\mathbf{\ell}+\mathbf{\delta}}^{-}+S_{A\mathbf{\ell}}^{-}%
S_{B\mathbf{\ell}+\mathbf{\delta}}^{+}\right)  \right]  ,\label{Haf}%
\end{equation}
where $\mathbf{\ell}$ runs over all sites of sublattice $A$. In this
paper, we only consider the contribution from the nearest neighbors.

The antiferromagnetic ground state is approximately taken as the
N\'{e}el state, in which the $z$ component of each spin is $S$ in
sublattice $A$, and $-S$ in sublattice $B$.

The excited state of antiferromagnets can be treated by the similar
treatment of ferromagnets. Similar to equation (\ref{amreal}),
introduce two sets of
operator realizations for sublattice $A$ and $B$, respectively:%
\begin{align}
S_{A\mathbf{\ell}}^{+}  &  =\sqrt{2S}\left(  1-\frac{N_{\mathbf{\ell}}}%
{4S}\right)  a_{\mathbf{\ell}},\nonumber\\
S_{A\mathbf{\ell}}^{-}  &
=\sqrt{2S}a_{\mathbf{\ell}}^{\dagger}\left(
1-\frac{N_{\mathbf{\ell}}}{4S}\right)  ,\nonumber\\
S_{A\mathbf{\ell}}^{z}  &  =S-N_{\mathbf{\ell}},\label{realA}%
\end{align}
and%
\begin{align}
S_{B\mathbf{\ell}}^{+}  &
=\sqrt{2S}c_{\mathbf{\ell}}^{\dagger}\left(
1-\frac{N_{\mathbf{\ell}}}{4S}\right)  ,\nonumber\\
S_{B\mathbf{\ell}}^{-}  &  =\sqrt{2S}\left(  1-\frac{N_{\mathbf{\ell}}}%
{4S}\right)  c_{\mathbf{\ell}},\nonumber\\
S_{B\mathbf{\ell}}^{z}  &  =-\left(  S-N_{\mathbf{\ell}}\right)
,\label{realB}%
\end{align}
where $a_{\mathbf{\ell}}^{\dagger}$, $a_{\mathbf{\ell}}$ and
$c_{\mathbf{\ell }}^{\dagger}$, $c_{\mathbf{\ell}}$ are the creation
and annihilation operators of the spin deviations on sublattice $A$
and $B$, satisfying the operator relations (\ref{icr}) and
(\ref{Nl}), respectively.

Substituting (\ref{realA}) and (\ref{realB}) into Hamiltonian
(\ref{Haf})
gives%
\begin{align}
H  &  =2J\sum_{\mathbf{\ell}\mathbf{\delta}}\left\{  \left(
S-N_{\mathbf{\ell
}}\right)  \left(  -S+N_{\mathbf{\ell}+\mathbf{\delta}}\right)  \right. \nonumber\\
&  \left.  +S\left[  \left(  1-\frac{N_{\mathbf{\ell}}}{4S}\right)
a_{\mathbf{\ell}}\left(  1-\frac{N_{\mathbf{\ell}+\mathbf{\delta}}}%
{4S}\right)
c_{\mathbf{\ell}+\mathbf{\delta}}+a_{\mathbf{\ell}}^{\dagger
}\left(  1-\frac{N_{\mathbf{\ell}}}{4S}\right)  c_{\mathbf{\ell}%
+\mathbf{\delta}}^{\dagger}\left(  1-\frac{N_{\mathbf{\ell}+\mathbf{\delta}}%
}{4S}\right)  \right]  \right\}  .
\end{align}
For low-lying excitations, we only take the contribution from the
terms
quadratic in the creation and annihilation operators into account:%
\begin{equation}
H\simeq-2JWZS^{2}+2JS\sum_{\mathbf{\ell}\mathbf{\delta}}\left(
N_{\mathbf{\ell}}+N_{\mathbf{\ell}+\mathbf{\delta}}\right)  +2JS\sum
_{\mathbf{\ell}\mathbf{\delta}}\left(
a_{\mathbf{\ell}}c_{\mathbf{\ell
}+\mathbf{\delta}}+a_{\mathbf{\ell}}^{\dagger}c_{\mathbf{\ell}+\mathbf{\delta
}}^{\dagger}\right)  .\label{Haprxmt}%
\end{equation}
Here $W$ denotes the number of the sites of sublattice $A$.

Introduce the transformations%
\begin{equation}
a_{\mathbf{\ell}}=\frac{1}{\sqrt{W}}\sum_{\mathbf{k}}e^{-i\mathbf{k}%
\cdot\mathbf{\ell}}b_{\mathbf{k}}\text{ and }a_{\mathbf{\ell}}^{\dagger}%
=\frac{1}{\sqrt{W}}\sum_{\mathbf{k}}e^{i\mathbf{k}\cdot\mathbf{\ell}%
}b_{\mathbf{k}}^{\dagger},\label{atobAF}%
\end{equation}
and%
\begin{equation}
c_{\mathbf{\ell}}=\frac{1}{\sqrt{W}}\sum_{\mathbf{k}}e^{-i\mathbf{k}%
\cdot\mathbf{\ell}}d_{\mathbf{k}}\text{ and }c_{\mathbf{\ell}}^{\dagger}%
=\frac{1}{\sqrt{W}}\sum_{\mathbf{k}}e^{i\mathbf{k}\cdot\mathbf{\ell}%
}d_{\mathbf{k}}^{\dagger}.\label{ctodAF}%
\end{equation}
Using the relation between $N_{\mathbf{\ell}}$ and $a_{\mathbf{\ell}}%
^{\dagger}$, $a_{\mathbf{\ell}}$, $c_{\mathbf{\ell}}^{\dagger}$, and
$c_{\mathbf{\ell}}$ and substituting the above transformations into
(\ref{Haprxmt}), up to the quadratic contribution, gives%
\begin{equation}
H\simeq-2JWZS^{2}+\frac{\left(  2n+1\right)  ^{2}}{2n\left(
n+1\right)
}JSZ\sum_{\mathbf{k}}\left(  b_{\mathbf{k}}^{\dagger}b_{\mathbf{k}%
}+d_{\mathbf{k}}^{\dagger}d_{\mathbf{k}}\right)  +2JSZ\sum_{\mathbf{k}}%
\gamma_{\mathbf{k}}\left(  b_{\mathbf{k}}d_{-\mathbf{k}}+b_{\mathbf{k}%
}^{\dagger}d_{-\mathbf{k}}^{\dagger}\right)  .\label{HaprxmtAF}%
\end{equation}

For diagonalizing the Hamiltonian (\ref{HaprxmtAF}), we introduce
the
Bogoliubov transformation which mixes the operators of the two sublattices:%
\begin{align}
b_{\mathbf{k}}  &  =u_{\mathbf{k}}\alpha_{\mathbf{k}}+v_{\mathbf{k}}%
\beta_{\mathbf{k}}^{\dagger},\text{ \ \ }d_{-\mathbf{k}}=u_{\mathbf{k}}%
\beta_{\mathbf{k}}+v_{\mathbf{k}}\alpha_{\mathbf{k}}^{\dagger},\nonumber\\
b_{\mathbf{k}}^{\dagger}  &
=u_{\mathbf{k}}\alpha_{\mathbf{k}}^{\dagger
}+v_{\mathbf{k}}\beta_{\mathbf{k}},\text{ \ \
}d_{-\mathbf{k}}^{\dagger
}=u_{\mathbf{k}}\beta_{\mathbf{k}}^{\dagger}+v_{\mathbf{k}}\alpha_{\mathbf{k}%
}.\label{BT}%
\end{align}
It can be checked directly that the Hamiltonian (\ref{HaprxmtAF})
can be diagonalized when $u_{\mathbf{k}}$ and $v_{\mathbf{k}}$ are
taken as
\begin{align}
u_{\mathbf{k}}^{2}  &  =\frac{1}{2}\left\{  \left[
1-\frac{16n^{2}\left( n+1\right)  ^{2}}{\left(  2n+1\right)
^{4}}\gamma_{\mathbf{k}}^{2}\right]
^{-1/2}+1\right\}  ,\nonumber\\
v_{\mathbf{k}}^{2}  &  =\frac{1}{2}\left\{  \left[
1-\frac{16n^{2}\left( n+1\right)  ^{2}}{\left(  2n+1\right)
^{4}}\gamma_{\mathbf{k}}^{2}\right]
^{-1/2}-1\right\}  .\label{uv}%
\end{align}
Then, from equation (\ref{HaprxmtAF}), ignoring the high-order
contribution,
we achieve%
\begin{align}
H  &  =-2JWZS^{2}+2JSZ\sum_{\mathbf{k}}\left\{  \left[  \frac{\left(
2n+1\right)  ^{2}}{4n\left(  n+1\right)  }u_{\mathbf{k}}^{2}+\gamma
_{\mathbf{k}}u_{\mathbf{k}}v_{\mathbf{k}}\right]  \left(  \alpha_{\mathbf{k}%
}^{\dagger}\alpha_{\mathbf{k}}+\beta_{\mathbf{k}}^{\dagger}\beta_{\mathbf{k}%
}\right)  \right. \nonumber\\
&  \left.  +\left[  \frac{\left(  2n+1\right)  ^{2}}{4n\left(
n+1\right)
}v_{\mathbf{k}}^{2}+\gamma_{\mathbf{k}}u_{\mathbf{k}}v_{\mathbf{k}}\right]
\left(  \alpha_{\mathbf{k}}\alpha_{\mathbf{k}}^{\dagger}+\beta_{\mathbf{k}%
}\beta_{\mathbf{k}}^{\dagger}\right)  \right\}  .\label{Hab}%
\end{align}
By the operator relation%
\begin{equation}
\alpha_{\mathbf{k}}^{\dagger}\alpha_{\mathbf{k}}=N_{\alpha\mathbf{k}}%
\frac{1-\frac{N_{\alpha\mathbf{k}}-1}{n}}{\left(  1-\frac{N_{\alpha\mathbf{k}%
}-1}{2n}\right)  ^{2}}\text{ \ \ and \ \ }\alpha_{\mathbf{k}}\alpha
_{\mathbf{k}}^{\dagger}=\left(  N_{\alpha\mathbf{k}}+1\right)  \frac
{1-\frac{N_{\alpha\mathbf{k}}}{n}}{\left(  1-\frac{N_{\alpha\mathbf{k}}}%
{2n}\right)  ^{2}}%
\end{equation}
(the operator relation for $\beta_{\mathbf{k}}$ is the same as that
of $\alpha_{\mathbf{k}}$), ignoring the high-order contribution,
from (\ref{Hab}), we achieve
\begin{equation}
H=-2JWZS^{2}+4JSZ\sum_{\mathbf{k}}\left[  \frac{\left(  2n+1\right)  ^{2}%
}{4n\left(  n+1\right)  }v_{\mathbf{k}}^{2}+\gamma_{\mathbf{k}}u_{\mathbf{k}%
}v_{\mathbf{k}}\right]
+\sum_{\mathbf{k}}\hbar\omega_{\mathbf{k}}\left(
N_{\alpha\mathbf{k}}+N_{\beta\mathbf{k}}\right)  ,
\end{equation}
where the dispersion relation for antiferromagnetic magnons is%
\begin{equation}
\hbar\omega_{\mathbf{k}}=2JSZ\left[  u_{\mathbf{k}}^{2}+v_{\mathbf{k}}%
^{2}+2\gamma_{\mathbf{k}}u_{\mathbf{k}}v_{\mathbf{k}}+\frac{1}{4n\left(
n+1\right)  }v_{\mathbf{k}}^{2}-\frac{1}{\left(  2n+1\right)  ^{2}}%
\gamma_{\mathbf{k}}u_{\mathbf{k}}v_{\mathbf{k}}\right]  .\label{spectrumAF}%
\end{equation}

Similar to the ferromagnetic case, the antiferromagnet magnon obeys
intermediate statistics rather than Bose-Einstein statistics. When
approximately regarding the magnon as bosons, based on the
Holstein-Primakoff
representation, the dispersion relation of the antiferromagnet magnon reads%

\begin{equation}
\hbar\omega_{\mathbf{k}}^{HP}=2JSZ\left(  u_{\mathbf{k}}^{HP2}+v_{\mathbf{k}%
}^{HP2}+2\gamma_{\mathbf{k}}u_{\mathbf{k}}^{HP}v_{\mathbf{k}}^{HP}\right)
,\label{HPAF}%
\end{equation}
where the superscript "$HP$" denotes that the corresponding result
comes from the method of the Holstein-Primakoff representation.
Comparing the dispersion relation (\ref{spectrumAF}) with
(\ref{HPAF}), we can see that when the maximum occupation number
$n\rightarrow\infty$, our result returns to the result of the
Holstein-Primakoff representation which regards magnons as bosons.
The magnitude of the modification relies on the value of $n$.

\section{Comparison with other schemes of intermediate
statistics\label{HWandanyon}}

\subsection{Comparison with Haldane-Wu fractional statistics}

The above result shows that the magnons obey a kind of intermediate
statistics with the statistical distribution (\ref{distribution}).
Let us compare this statistical distribution with another kind of
intermediate statistics ----- Haldane-Wu fractional statistics
\cite{gPauli}. The Haldane-Wu
distribution function reads%
\begin{equation}
\left\langle N_{\mathbf{k}}^{HW}\right\rangle
=\frac{1}{\omega^{-1}+\left( g-1\right)  },
\end{equation}
where $\omega$ is determined by $g\ln\left(  1-\omega\right)
-\ln\omega =\beta\left(  \varepsilon_{\mathbf{k}}-\mu\right)  $.

In the Haldane-Wu distribution, there is an intermediate-statistics
parameter $g$, and in the intermediate statistics obeyed by magnons,
the intermediate-statistics parameter is $n$. A relation between
these two intermediate-statistics parameters can be obtained by
comparing the second virial coefficients. The second virial
coefficient of a $\nu$-dimensional ideal magnon gas with the
dispersion relation $\varepsilon\propto p^{s}$ can
be obtained directly:%
\begin{equation}
a_{2}=\displaystyle-\frac{1}{2^{\nu/s+1}}\frac{\nu\Gamma(\frac{\nu}{2}%
)}{2\Gamma\left(  \frac{\nu}{s}+1\right)  }\frac{4n^{3}+8n^{2}-2}%
{4n^{3}+8n^{2}+2n-1}.\label{svc}%
\end{equation}
The second virial coefficient of an ideal gas obeying Haldane-Wu
fractional statistics reads $a_{2}^{HW}=-\left(  1-2g\right)
/2^{\nu/s+1}$ \cite{Kha}.
Comparing these two second virial coefficients gives%
\begin{equation}
g=\displaystyle\frac{1}{2}\left[  1-\frac{\nu\Gamma(\frac{\nu}{2})}%
{2\Gamma\left(  \frac{\nu}{s}+1\right)  }\frac{4n^{3}+8n^{2}-2}{4n^{3}%
+8n^{2}+2n-1}\right]  .
\end{equation}

\subsection{Comparison with the fractional statistics of anyons}

It is also interesting to compare this intermediate statistics with
the fractional statistics of anyons, another scheme of intermediate
statistics \cite{Wilczek}. For the case of anyon, we of course only
focus on two dimensions.

The concept of anyons is introduced by generalizing the change of
the phase factor of a wavefunction when two identical particles
exchange to an arbitrary phase factor $e^{i\theta}$. $\theta=0$ and
$\theta=\pi$ correspond to Bose-Einstein and Fermi-Dirac cases,
respectively.

The second virial coefficient of an anyon gas reads \cite{Kha,CGO}%
\begin{equation}
a_{2}=-\frac{1}{4}\left(  1-4\alpha+2\alpha^{2}\right)  ,
\end{equation}
where $\alpha=\theta/\pi$. Comparing this result with the second
virial
coefficient (\ref{svc}) with $\nu=2$ and $s=2$ gives%
\begin{equation}
\alpha=1-\sqrt{1-\frac{2n+1}{8n^{3}+16n^{2}+4n-2}}.
\end{equation}

\section{Comparison with the exact result of the Bethe ansatz method: the
spectrum\label{Bethe}}

In this section, we compare our result with the exact solution of
the one-dimensional spin $1/2$ Heisenberg model.

By the Bethe ansatz, one can find the exact solutions of certain
one-dimensional quantum many-body models. Taking ferromagnets as an
example, we compare our result given in section \ref{FSW} with the
exactly solved one-dimensional $s=1/2$ Heisenberg model with two
down spins.

\begin{figure}[ptb]
\begin{center}
\includegraphics[height=7in,width=4in]
{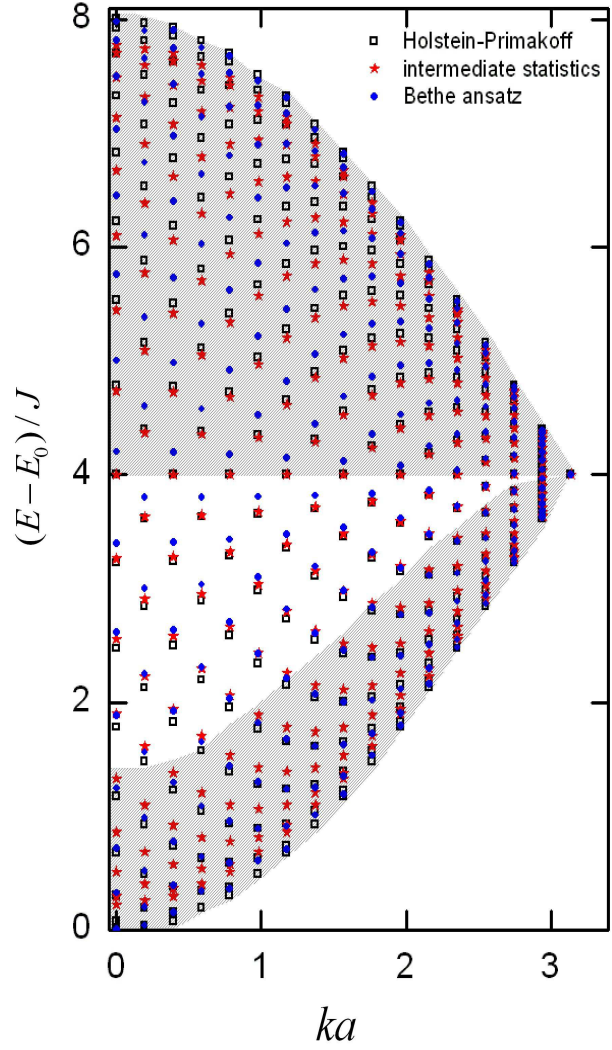}
\end{center}
\caption{The spectra of the one-dimensinal $s=1/2$ Heisenberg chain
given by the Bethe ansatz method (the exact solution), by the
Holstein-Primakoff method, and by the intermediate-statistics
method. In the unshadowed area, in comparison with the exact result,
the intermediate-statistics result is better
than the result given by the Holstein-Primakoff method.}%
\end{figure}

The exact spectrum of the one-dimensional $s=1/2$ Heisenberg model
with two
down spins is given by \cite{Bethe,Cini,Solyom}%
\begin{equation}
E-E_{0}=J\left(  2-\cos k_{1}a-\cos k_{2}a\right)  ,
\end{equation}
where $a$ is the lattice constant and $k_{1}$, $k_{2}$ are determined by%
\begin{align}
Nk_{1}a &  =2\pi\lambda_{1}+\theta,\nonumber\\
Nk_{2}a &  =2\pi\lambda_{2}-\theta,\nonumber\\
2\cot\frac{\theta}{2} &  =\cot\frac{k_{1}a}{2}-\cot\frac{k_{2}a}{2},
\end{align}
with $\lambda_{1}$, $\lambda_{2}=0,1,2,\cdots,N-1$ and
$\lambda_{2}\geq \lambda_{1}$.

Moreover, our result of the spectrum for the corresponding case can
be directly obtained by equation (\ref{spectrum2}) with $n=1$.

The spectra obtained by the Bethe ansatz (the exact one), by the
Holstein-Primakoff method, and by the intermediate-statistics method
are sketched in figure 1.

Comparing with the exact result obtained by the Bethe ansatz, we can
see that in some cases (the unshadowed area) our result (the
intermediate-statistics magnons) is more accurate than the standard
Holstein-Primakoff result (the bosonic magnons).

\section{Comparing with the experiment\label{numberical}}

From equation (\ref{all}), we can obtain the relation between the
spin-wave energies and the temperature by the self-consistent
calculation. We will consider the spin-wave dispersion relation of
$EuO$ since the spin-wave dispersion for $EuO$ is isotropic. The
$Eu^{2+}$ ions in $EuO$ form simple fcc lattices, so the number of
the nearest neighbors and next-to-nearest neighbors are $Z_{1}=12$
and $Z_{2}=6$, the exchange parameters $J_{1}$ to nearest neighbors
and $J_{2}$ to next-to-nearest neighbors are $J_{1}=0.606k_{B}$ and
$J_{2}=0.119k_{B}$ \cite{DA3}, where $k_{B}$ is the Boltzmann
constant, and $S=7/2$.

\begin{figure}[ptb]
\begin{center}
\includegraphics[height=5in,width=7.3in]
{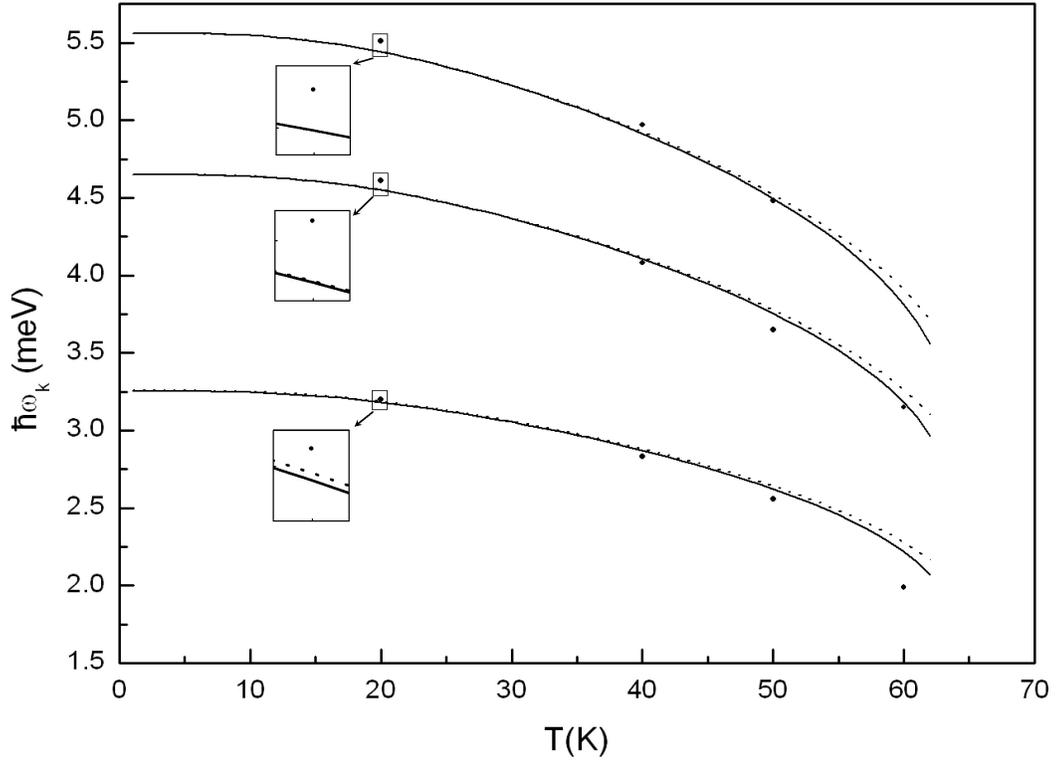}
\end{center}
\caption{The spin-wave energies in EuO. The dotted lines represent
the intermediate-statistics result and the solid lines represent the
result of the Holstein-Primakoff representation. The experimental
data are taken from
\cite{GMP}.}%
\end{figure}

The calculation results are plotted in figure 2. The experimental
data are taken from \cite{GMP}.

In comparison with the experimental data, we can see that at low
temperatures and low frequencies, the result of the
intermediate-statistics spin waves is more accurate than the result
of the bosonic spin waves, and at high temperatures and high
frequencies, the result of bosonic spin waves is better.

\section{Discussion and Conclusions \label{conclusions}}

It is shown that magnons, the elementary excitation of a Heisenberg
magnetic system, obey a kind of intermediate statistics with a
maximum occupation number $n=2S$. In the common treatment, the
solution of the spectrum of a magnetic system is based on the
Holstein-Primakoff representation which is a bosonic operator
relation with an additional restriction on the occupation number.
Since the information of the maximum occupation number is not
embodied in the operator relation, the influence of the restriction
on the occupation number is not reflected in the result of the
spectrum. Consequently, the magnons are approximately treated as
bosons in the Holstein-Primakoff treatment: the commutation relation
of creation and annihilation operators is taken as the bosonic
commutation relation and the statistical distribution is taken as
the Bose-Einstein distribution.

In this paper, we construct an intermediate-statistics operator
realization in which the information of the maximum occupation
number which is equal to an integer $n$ is embodied in the
commutation relation of creation and annihilation operators rather
than putting in a restriction on the occupation number by hand.
Then, from the operator relations, we directly deduce the
corresponding statistical distribution function, which is a modified
Bose-Einstein statistical distribution and will return to the
Bose-Einstein distribution when taking the maximum occupation number
$n$ to be $\infty$.

It is the starting point that there is a natural relation between
the angular momentum and the intermediate statistics with a given
maximum occupation number. For the intermediate statistics with a
maximum occupation number $n$, there are $n+1$ states, $\left\vert
0\right\rangle $, $\left\vert 1\right\rangle $, $\left\vert
2\right\rangle $,$\cdots$, $\left\vert n\right\rangle $. For the
angular momentum $S$, there are $2S+1$ states $\left\vert
-S\right\rangle $, $\left\vert -S+1\right\rangle $,$\cdots$,
$\left\vert S-1\right\rangle $, $\left\vert S\right\rangle $. This
naturally leads us to relate the $n+1$ states, $\left\vert
0\right\rangle $, $\left\vert 1\right\rangle $,$\cdots$, $\left\vert
n\right\rangle $, to the $2S+1$ angular momentum states, $\left\vert
-S\right\rangle $, $\cdots$, $\left\vert S\right\rangle $.
Consequently, we have the relation $n+1=2S+1$, and then $n=2S$. From
this, we can construct an intermediate-statistics realization and
reveal that the statistics of magnons is intermediate statistics.

Based on the results of the intermediate statistics provided in
sections \ref{operator} and \ref{distributionfunction}, we calculate
the dispersion relation of the ferromagnetic spin wave up to the
quartic contribution, in which the influence of the interaction
between magnons is taken into account, and the dispersion relation
of the antiferromagnetic spin wave up to the quadratic contribution.
Compared to the result of the Holstein-Primakoff representation, the
bosonic operator relation is replaced by the intermediate-statistics
operator relation, and the Bose-Einstein distribution is replaced by
the intermediate-statistics distribution, so the influence of the
restriction on the occupation number is naturally taken into
account. Moreover, we also take into account the
next-to-nearest-neighbor contribution and the influence of the
classical dipolar interaction which is caused by the interaction
between the magnetic moments in the ferromagnetic case.

Magnons obey a kind of intermediate statistics. As comparisons, we
discuss the relations among the intermediate statistics obeyed by
magnons, Haldane-Wu fractional statistics, and the fractional
statistics of anyons. The relations among the three
intermediate-statistics parameters are given.

For discussing the validity of our result, we compare our result
with the exact solution of the one-dimensional spin $1/2$ Heisenberg
model obtained by the Bethe-ansatz method.

Our results of the dispersion relation of the magnetic systems are
based on intermediate statistics, in which the maximum occupation
number is an integer $n$ equaling $2S$. We compare our result with
the result by the Holstein-Primakoff representation in which magnons
are assumed to obey Bose-Einstein statistics and with the
experimental data of $EuO$. The result compares well with the
experiment.

In a word, the elementary excitation of the Heisenberg magnetic
system obeys a kind of intermediate statistics with an finite
maximum occupation number $n=2S$ rather than Bose-Einstein
statistics.

\vskip 0.5cm \noindent\textbf{Acknowledgements} We are very indebted
to Dr. G. Zeitrauman for his encouragement. This work is supported
in part by NSF of China under Grant No. 10605013 and the Hi-Tech
Research and Development Programme of China under Grant No.
2006AA03Z407.

\end{document}